\documentstyle[psfig,12pt]{article}

 \newfont{\titlefont}{cmssbx10 scaled\magstep5}

 \newcommand{\CL}{{\cal L}}

 \newcommand{\CO}{{\cal O}}

 \newcommand{\bea}{\begin{eqnarray}}  \newcommand{\eea}{\end{eqnarray}}
 \newcommand{\beq}{\begin{equation}}  \newcommand{\eeq}{\end{equation}}
 \newcommand{\non}{\nonumber}  
   
 \newcommand{\lmk}{\left(}  \newcommand{\rmk}{\right)}
 \newcommand{\lkk}{\left[}  \newcommand{\rkk}{\right]}
 \newcommand{\lhk}{\left \{ }  \newcommand{\rhk}{\right \} }
 \newcommand{\del}{\partial}  
 \newcommand{\vect}[1]{\mbox{\boldmath${#1}$}}
 \newcommand{\bib}{\bibitem} \newcommand{\new}{\newblock}
 \newcommand{\la}{\left\langle} \newcommand{\ra}{\right\rangle}

\begin{document}

\begin{flushright}
  UTAP-290 \\ YITP-98-48 \\ SU-ITP-98/51 \\ ICRR-Report-429-98-25
\end{flushright}

\begin{center}
  {\Large \bf Numerical analysis of formation and evolution of global
    strings in $2+1$ dimensions \\}
  \vskip 1cm
  {\large Masahide Yamaguchi} \\
  \vskip 0.2cm
  {\large \em Department of Physics, University of Tokyo, \\
    Tokyo, 113-0033, Japan}
  \vskip 0.5cm {\large Jun'ichi Yokoyama} \\
  \vskip 0.2cm {\large \em Department of Physics, Stanford University,
    \\ Stanford, CA94305-4060, USA \\
    and Yukawa Institute
    for Theoretical Physics, Kyoto University, \\
    Kyoto, 606-8502, Japan}
  \vskip 0.5cm {\large M. Kawasaki} \\
  \vskip 0.2cm {\large \em Institute for Cosmic Ray Research,
    University of Tokyo, \\
    Tokyo, 188-8502, Japan}
  \end{center}


\begin{abstract}
  We simulate the formation and the evolution of global strings taking
  into account the expansion of the universe and the concomitant
  change of the effective potential, that is, the change from the
  restoration stage of the global {\it U}(1)-symmetry to the broken
  stage.  Starting from the thermal equilibrium state, we run
  simulations for 100 sets of different initial conditions in 2+1
  dimensions. As a result, we find deviation from scale invariance.
  This is because the energy loss mechanism is pair annihilation
  rather than intercommutation, so that a string and an anti-string
  attract each other under a logarithmic potential.
\end{abstract}

\thispagestyle{empty} \setcounter{page}{0} \newpage
\setcounter{page}{1}


\section{Introduction}

\label{sec:introduction}

Since the possibility of the restoration of broken symmetries in the
very early universe was suggested \cite{KL}, the cosmological
consequences of the accompanying phase transitions have been
investigated by a number of authors \cite{phase}. As Kibble first
pointed out, topological defects of various types may be formed,
depending on the pattern of the symmetry breaking \cite{Kib} (see also
Ref.~\cite{defect} for review). If the symmetry group $G$ breaks into
a subgroup $H$ and the coset space $G/H$ is not simply-connected,
strings are formed. Strings as local defects have been extensively
investigated in the context of structure formation theory \cite{Bra}.
On the other hand, global strings are predicted in the axion theory of
the strong {\it CP} problem \cite{PQ}, and they may have radiated
significant amount of axions as the universe cools down from the
Peccei-Quinn breaking scale to the energy scale of QCD \cite{davis}.
Also, in condensed-matter systems, analogous vortices are
observed \cite{Zur}.

Most studies of the formation of strings have employed Monte-Carlo
simulations. Vachaspati and Vilenkin \cite{VV} first examined the
initial configuration of a string network. They found that it has the
statistical properties of a Brownian random walk (see also
Ref.~\cite{SF}), and found that a string network is made of closed
loops with a scale-invariant size distribution and infinite strings
occupying $80\%$ of the total string length. Since that time, many
authors have reexamined the problem in more realistic situations than
the original, focusing especially on the existence and the fraction
(if any) of infinite strings \cite{form}. However, as seen from the
fact that the above mentioned studies were carried out exclusively by
the Monte-Carlo simulations, little work on the dynamical aspects of
string formation has been done.

One exception is Borrill's paper \cite{Bor}, where the simulation of
the bubble nucleation process in a first order phase transition is
performed using a lattice-free method. However, in his approach the
subsequent evolution of the string network could not be followed. In
Ref.~\cite{Hod} the Langevin-equation approach is considered. But in
that study the effect of the expansion of the universe was considered
only partially. This is quiet different from the real dynamics. Ye and
Brandenberger \cite{YE}, on the other hand, performed a numerical
simulation properly taking cosmic expansion into account. But there, a
Mexican-hat potential was adopted from the outset, and the simulation
domain was so small that the whole system fell within one horizon
before they could conclude that the network had reached the scaling
regime. The evolution of local (gauged) strings has been investigated
by using the Nambu-Goto action. There are many analytical \cite{ana}
and numerical \cite{num} approaches, and it has been confirmed that
the string networks relax to the scaling solution by intercommutation
which generates closed loops that eventually disappear by radiating
gravitational waves or particles.  On the other hand, as for global
strings, the Kalb-Ramond action \cite{KR} must be used instead of the
Nambu-goto action in order to examine the evolution. But in that
approach it is difficult to include the long-range force between
infinite strings.  Here, we use an approach based on a complex scalar
field.

In this paper, we simulate both the formation and evolution of the
global string network by solving the dynamics of the complex scalar
field directly. Starting from a symmetry-restored stage, we follow the
dynamics of the complex scalar field, fully considering the expansion
of the universe and the associated change of the effective potential,
which exhibits spontaneous symmetry breaking below the critical
temperature.

The simulation is made in 2+1 dimensions as a first step to analyze
the full 3+1-dimensional analysis, because the spatial resolution and
total cosmic time are limited due to our limited computing power. Note
that in 2+1 dimensions strings are point-like, and each such string
corresponds to an infinite string in 3+1 dimensions so that the energy
loss mechanism is pair-annihilation rather than loop-formation through
intercommutation.  The direct emission of goldstone modes from strings
is suppressed because of the symmetry of the system. In a forth-coming
paper we will report the results of our simulations in 3+1 dimensions
and their cosmological implications \cite{YKY}.

The remainder of this paper is organized as follows. In the next
section, we establish the formulation for the numerical simulation. In
\S \ref{sec:results}, we give the result of the numerical simulation
and judge whether during the evolution the scaling behavior is
observed as expected from the analogy of local strings. In \S
\ref{sec:model}, the results obtained from the simulation are
explained analytically using a phenomenological model. Finally, \S
\ref{sec:summary} is devoted to summary and discussion.

\section{Formulation}

\label{sec:formulation}

We consider the following Lagrangian density for a complex scalar
field $\Phi(x)$: 
\beq
  \CL[\Phi] = \del_{\mu}\Phi \del^{\mu}\Phi^{\dagger} - V[\Phi] \:.
\eeq

\noindent
Here, the potential $V[\Phi]$ is given by
\beq
  V[\Phi] = \frac{\lambda}{2}\,(\Phi\Phi^{\dagger} - \eta^2)^2 \:.
\eeq

\noindent
Then the one-loop finite temperature effective potential $V_{\rm
eff}[\Phi]$ in the high temperature limit is written as
\beq
  V_{\rm eff}[\Phi] = \frac{\lambda}{2}(\Phi\Phi^{\dagger} - \eta^2)^2 
                 + \frac{\lambda}{3}T^2\,\Phi\Phi^{\dagger} \:.
\eeq

\begin{figure}[htb]
  \begin{center}
    \leavevmode\psfig{figure=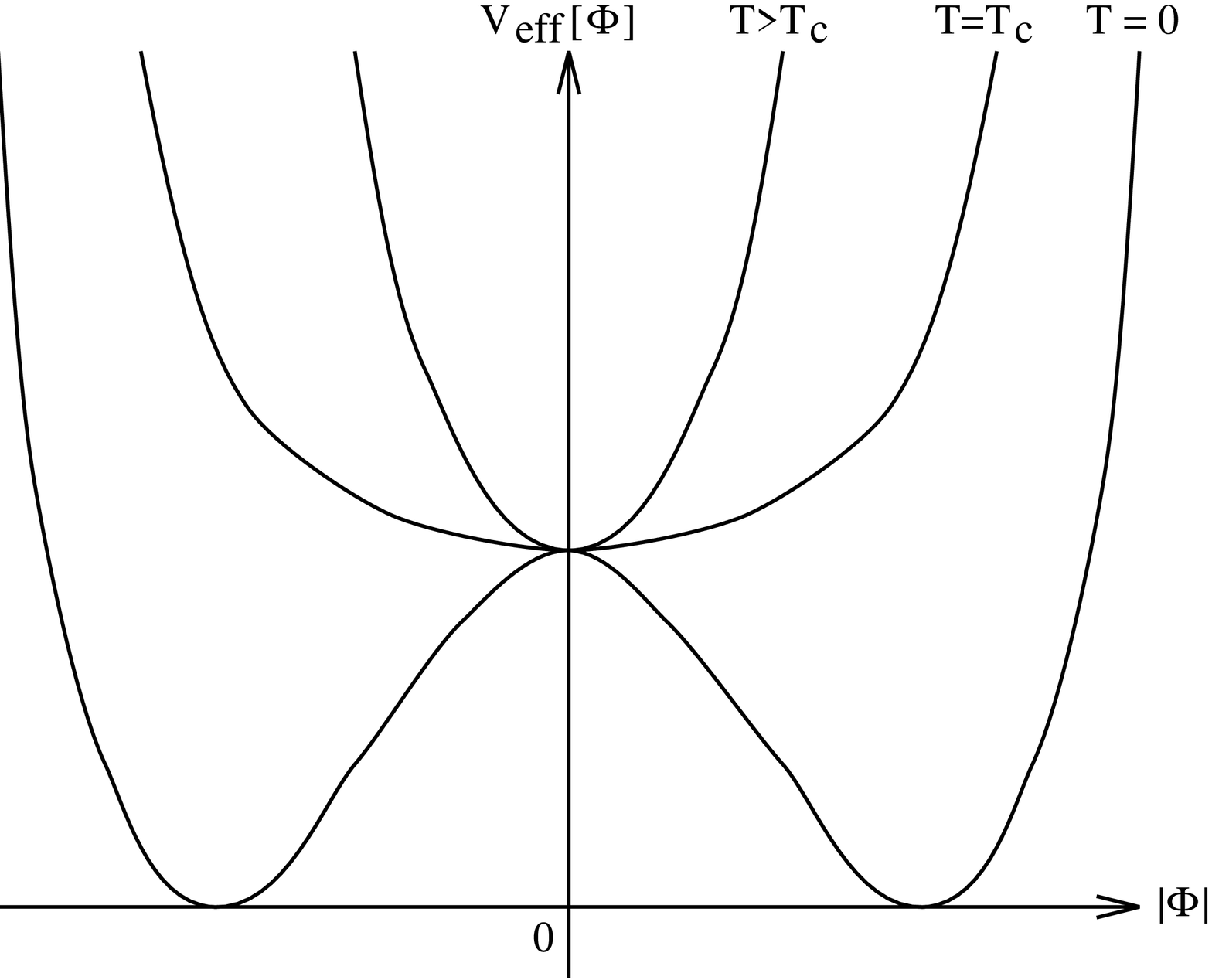,width=15cm}
  \end{center}
  \caption{One-loop finite temperature effective potential $V_{\rm
      eff}[\Phi]$ of the complex scalar field.}
  \label{fig:potential}
\end{figure}

\noindent
For $T > T_{c} = \sqrt{3}\eta$, the potential $V_{\rm eff}$ has a
minimum at $\Phi = 0$, and the $U(1)$ symmetry is restored. On the
other hand, new minima $|\Phi|_{min} = \eta\sqrt{1-(T/T_{c})^2}$
appear and the symmetry is broken for $T < T_{c}$ (Fig.\ 
\ref{fig:potential}). In this case the phase transition is of second
order.

In the expanding universe the effective Lagrangian density $\CL_{\rm
  eff}[\Phi]$ is given by
\beq
  \CL_{\rm eff}[\Phi] = g_{\mu\nu}(\del^{\mu}\Phi)(\del^{\nu}\Phi)^{\dagger}
                 - V_{\rm eff}[\Phi] \:,
\eeq 

\noindent
where $g_{\mu\nu}$ is identified with the Robertson-Walker metric.
Then the equation of motion is
\beq
  \ddot{\Phi}(x) + 3H\dot{\Phi}(x) - \frac{1}{a(t)^2}\nabla^2\Phi(x)
   = - V'_{\rm eff}[\Phi] \:,
\eeq

\noindent
where the prime represents the derivative $\del/\del\Phi^{\dagger}$,
and $a(t)$ is a scale factor.  The Hubble parameter $H = \dot
a(t)/a(t)$ and the cosmic time $t$ are given by
\bea
  H^2 = \frac{8\pi}{3 M_{\rm pl}^2} \frac{\pi^2}{30} g_{*} T^4 \:,
   ~~~~~
  t = \frac{1}{2H} = \frac{\xi}{T^2} \:,
\eea

\noindent
where $M_{\rm pl}$ is the Plank mass, $g_{*}$ is the total number of
degrees of freedom for the relativistic particles, and radiation
domination is assumed. We define the dimensionless parameter $\zeta$
as
\beq
  \zeta \equiv \frac{\xi}{\eta}  = \lmk \frac{45M_{\rm pl}^2}
      {16\pi^3g_{*} \eta^2}
  \rmk^{1/2} \:.
\eeq

\noindent
In our simulation, we take $\zeta = 10$, which corresponds to $\eta
\sim (10^{15}-10^{16})$ GeV, but the essential result is independent
of this choice. The energy density of each point is written as
\beq
  \rho(x) = \dot\Phi(x) \dot\Phi^{\dagger}(x) 
           + \frac{1}{a(t)^2}\nabla\Phi(x) \cdot \nabla\Phi^{\dagger}(x) 
            + V_{\rm eff}[\Phi]  \:.
\eeq

We simulate the system in 2+1 dimensions from the initial time
$t_{i} = t_{c}/4$, that is, $T_{i} = 2\,T_{c}$, to the final
time $t_{f} = 150\,t_{i} = 37.5\,t_{c}$, when $T_{f} \sim T_{c} /
6.1$. Since the $U(1)$ symmetry is restored at the initial time $t =
t_{i}$, we adopt as the initial condition the thermal equilibrium
state with the mass squared,
\beq
  m^2 = \left. \frac{d^2 V_{\rm eff}[|\Phi|]}{d|\Phi|^2} 
        \right|_{|\Phi|=0} \:,
\eeq

\noindent
which is the inverse curvature of the potential at the origin at $t =
t_{i}$. In the thermal equilibrium state, $\Phi$ and $\dot\Phi$ are
Gaussian distributed with the correlation functions,
\bea
  \la \beta|\Phi(\vect x)\Phi^{\dagger}(\vect y)|\beta
                             \ra_{\rm equal\hbox{-}time} &=&
   \int \frac{d\vect k}{(2\pi)^3} \frac1{2\sqrt{\vect k^2 + m^2}}
           \coth{\frac{\beta\sqrt{\vect k^2 + m^2}}{2}}
             e^{i\vect k \cdot (\vect x-\vect y)} \:, \non  \\
             && \\
  \la \beta|\dot\Phi(\vect x)\dot\Phi^{\dagger}(\vect y)|\beta
                             \ra_{\rm equal\hbox{-}time} &=&
   \int \frac{d\vect k}{(2\pi)^3} \frac{\sqrt{\vect k^2 + m^2}}{2}
           \coth{\frac{\beta\sqrt{\vect k^2 + m^2}}{2}}
             e^{i\vect k \cdot (\vect x-\vect y)} \:. \non \\
             &&
\eea

\noindent 
The functions $\Phi(\vect x)$ and $\dot\Phi(\vect y)$ are uncorrelated
for $\vect x \ne \vect y$. We generate these fields for the initial
condition in the momentum space, because the corresponding fields
$\tilde{\Phi}(\vect k)$ and $\tilde{\dot\Phi}(\vect k)$ are
uncorrelated there.

Hereafter we measure the scalar field in units of $t_{i}^{-1}$, $t$
and $x$ in units of $t_{i}$, and the energy density in units of
$t_{i}^{-4}$. The equation of motion and the total energy density are
given by
\bea
  && \hspace{-1.8cm} 
   \ddot{\Phi}(x) + \frac{3}{2t}\dot{\Phi}(x) - \frac{1}{t}\nabla^2\Phi(x)
   = - \lmk |\Phi|^2 + \frac{25}{9t} 
                                - \frac{25}{36} \rmk 
    \Phi^{\dagger} \:, \\
  && \hspace{-1.8cm} 
   \rho(x) = \dot\Phi(x) \dot\Phi^{\dagger}(x) 
           + \frac{1}{t}\nabla\Phi(x) \cdot \nabla\Phi^{\dagger}(x) 
            + \frac12 \lmk |\Phi|^2 - \frac{25}{36} \rmk^2
             + \frac{25}{9t} |\Phi|^2 \:,
\eea

\noindent
where $\lambda$ is set to unity.\footnote{Strictly speaking, the high
  temperature approximation is valid only when $\lambda$ is much
  smaller than unity. However, a change of $\lambda$ only scales the
  physical quantity and is not relevant to the final result.} The
scale factor $a(t)$ is normalized as $a(1) = 1$. We arrange $512
\times 512$ lattice points for our simulations. The lattice spacing
$\delta x$ is set to 0.1, and its physical length $\delta
x_{\rm{phys}}$ grows as $a(t)\delta x = 0.1\,t^{1/2}$. The time step
$\delta t$ is chosen to be $0.01$. At sufficiently low temperatures,
the typical width of the string $d$ is nearly $1/\eta$.  Then,
\beq
\hspace{-0.4 cm}
  \frac{H^{-1}}{\delta x_{\rm{phys}}} =
         20\,t^{1/2}, \qquad
         (\sim 245 \:\: {\rm at} \:\: t = t_{f})
\eeq

\vspace{0.2 cm}
\beq
  \frac{d}{\delta x_{\rm{phys}}} = 
         12\,t^{-1/2}.  \qquad
         (\sim 0.98 \:\: {\rm at} \:\: t = t_{f})
\eeq

\noindent
Even at the final time $t_{f}$, the resolution of the simulation is
sufficiently good, and the simulation box is larger than the horizon
scale, so that we can safely impose a periodic boundary condition.

Under these circumstances, we simulated the system using the second
order leap-frog method and the Crank-Nicholson scheme over 100 runs
with a different random realization of the initial conditions.

\section{Results}

\label{sec:results}

\begin{figure}[htb]
  \begin{center}
    \leavevmode\psfig{figure=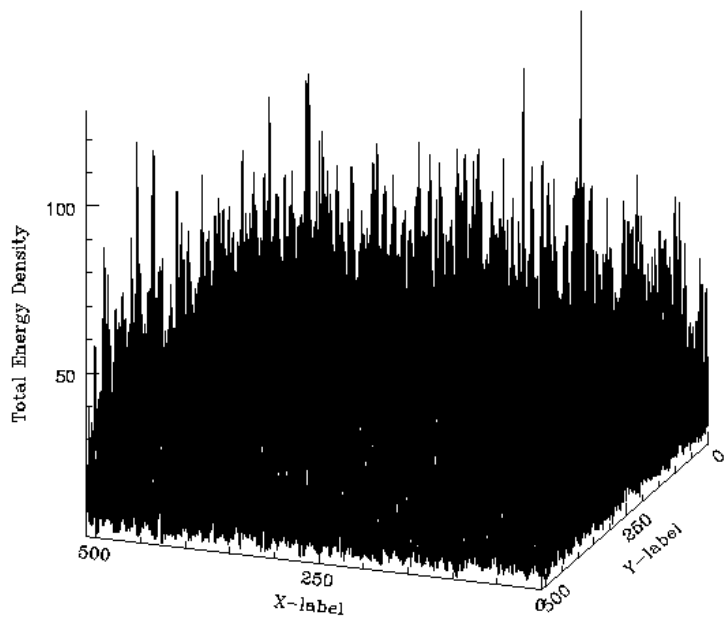,height=15cm}
  \end{center}
  \caption{The total energy density for each lattice point is
    depicted at $t = t_{i} = 1$.}
  \label{fig:evolution1}
\end{figure}

\begin{figure}[htb]
  \begin{center}
    \leavevmode\psfig{figure=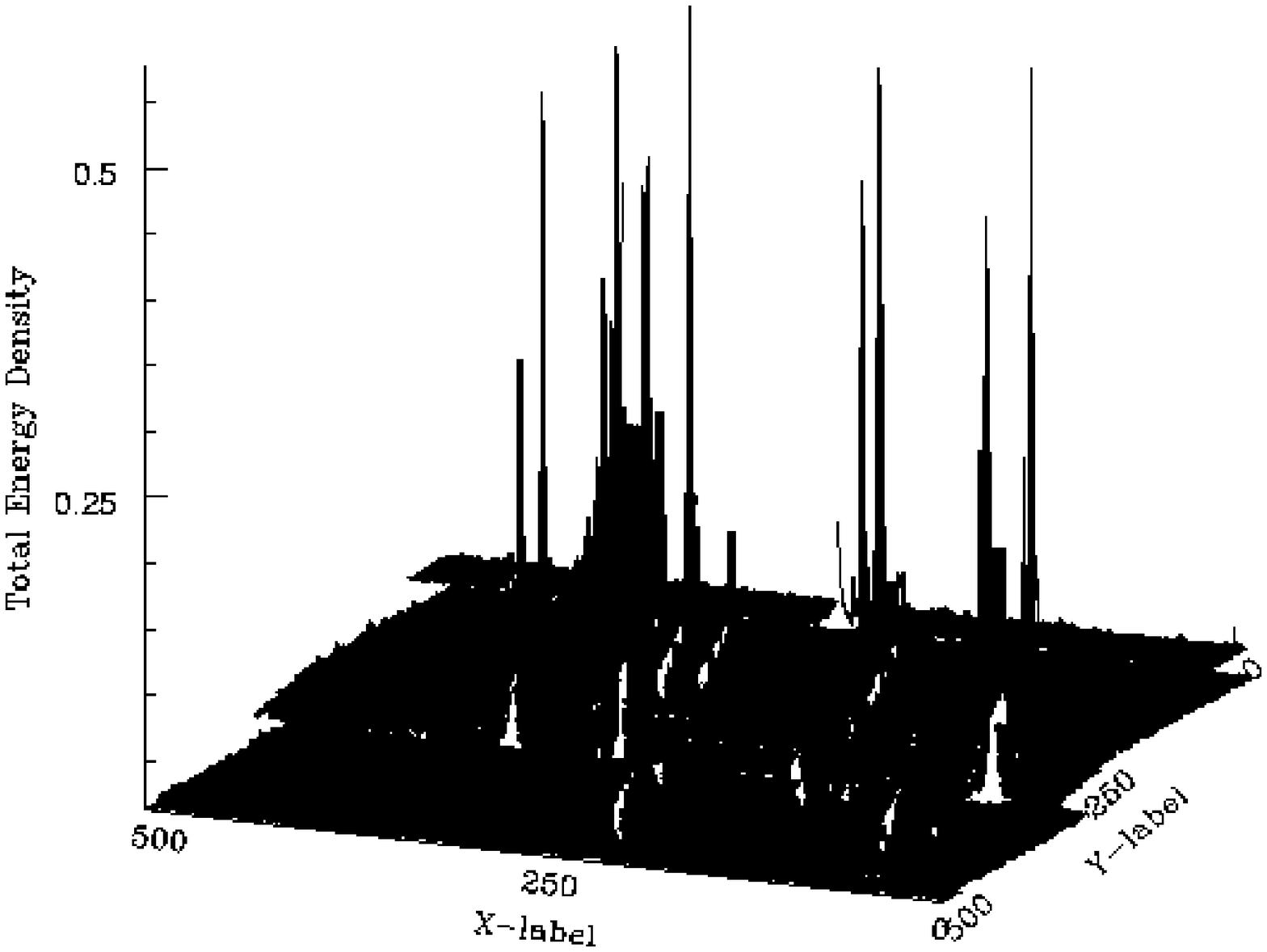,height=15cm}
  \end{center}
  \caption{That at $t = 150$.}
  \label{fig:evolution4}
\end{figure}

First we depict in Figs.\ 
\ref{fig:evolution1}$\sim$\ref{fig:evolution4} how strings are
generated as the symmetry is spontaneously broken and the gradient and
kinetic energy density decreases.  In these figures, the total energy
density at each lattice point is depicted at $t = 1, 50, 100, 150$.
One can see that initially the field fluctuates very strongly and the
gradient and kinetic energy dominates over the potential energy. But
as the universe expands and the temperature decreases, the false
vacuum energy becomes dominant, and the string configurations become
stable.

Next, in order to judge whether the string network really enters a
scaling regime in this case, we count the number of strings per
horizon volume at various times. Since spacetime is discretized in our
simulations, a point at which $\Phi = 0$ corresponding to a string
core is not necessarily situated at a lattice point. In the worst
case, a point at which $\Phi = 0$ lies at the center of a plaquette.
Therefore in order to identify a string, we stipulate that a lattice
is identified with a part of a string if the potential energy density
there is larger than that corresponding to the field value of a static
string solution at $r = \delta x_{\rm{phys}}/\sqrt{2}$,\footnote{In
  fact, the potential energy is peaked so highly at the string that
  the identification of the string depends very little on this
  criterion.} which is obtained by solving the equation
\beq
  \frac{\del^2\rho}{\del r^2} + \frac{1}{r}\frac{\del\rho}{\del r}
   - \frac{\rho}{r^2} - V'_{\rm eff}[\rho] = 0 \:,
\eeq

\noindent
with $\Phi(r, \theta) \equiv \rho(r)e^{i\theta}$. Then we regard a
connected region of lattices satisfying the above condition as one
string. Around $t = 20$, the kinetic and the gradient energy is
sufficiently suppressed in comparison with the vacuum energy of the
string so that we can identify the string unambiguously. We carried
out this identification procedure every 10 steps after $t = 20$.

\begin{figure}[htb]
  \begin{center}
    \leavevmode\psfig{figure=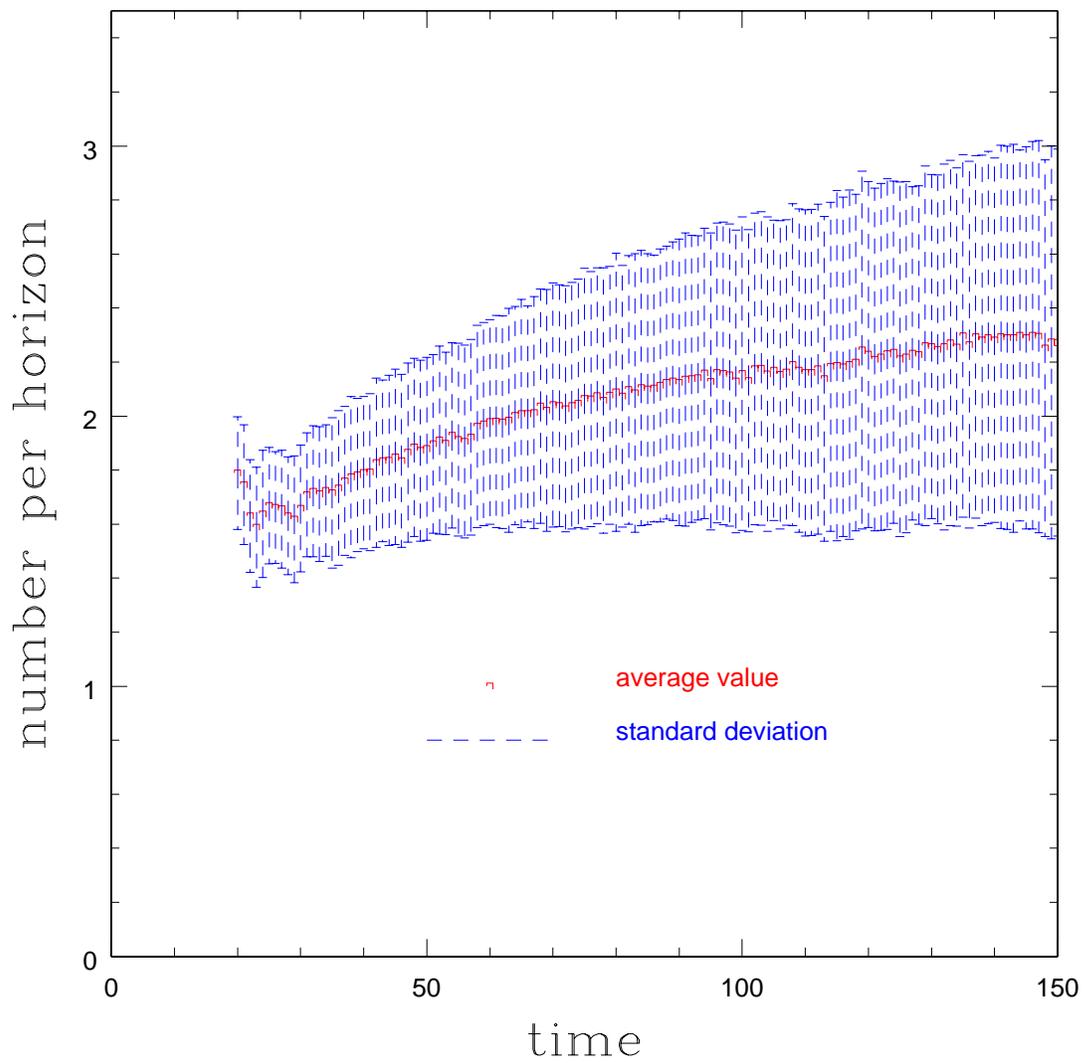,width=15cm}
  \end{center}
  \caption{The number of strings per horizon volume is depicted. Dots
    represent the average value over 100 realizations.
    Error-bars the standard deviation.}
  \label{fig:simulation}
\end{figure}

\begin{figure}[htb]
  \begin{center}
    \leavevmode\psfig{figure=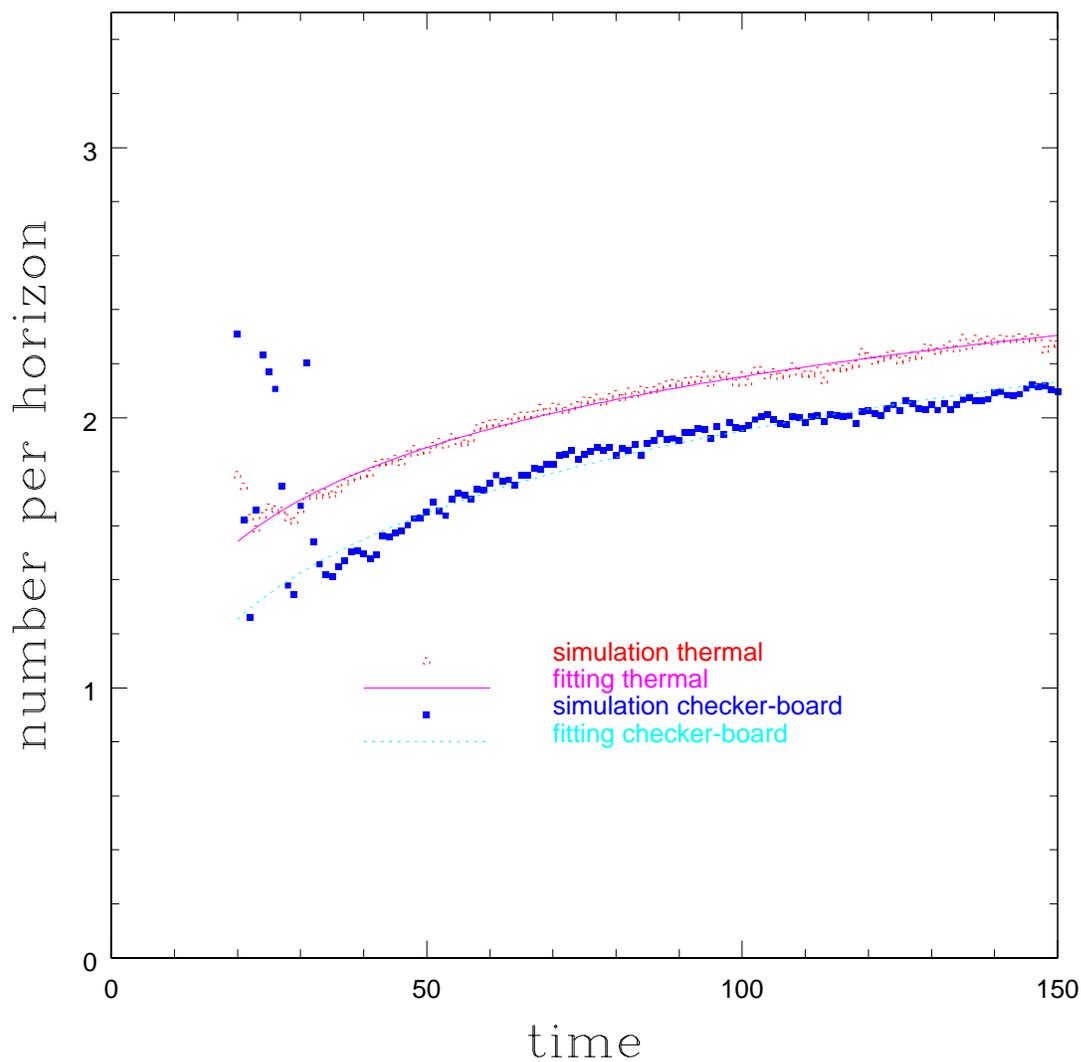,width=15cm}
  \end{center}
  \caption{Open dots represent the number of strings per horizon volume
    averaged over those from 100 initial conditions which are the
    thermal equilibrium states. On the other hand, filled dots from
    checker-board initial conditions. Also, a solid line represents
    the fitting formula for simulation results from thermal initial
    conditions and a dotted line from checker-board ones.}
  \label{fig:fitting}
\end{figure}

Figure\ \ref{fig:simulation} displays that time evolution of the
number of strings per horizon volume averaged over 100 realizations.
It is clear that the number is not constant but gradually increases
with the cosmic time. Since the rate of increase is a decreasing
function of time, one may think that we have observed only a
relaxation stage and that it will eventually approach a constant.  In
order to judge if this is the case, we run a different set of
simulations with different initial conditions, in which strings and
anti-strings are arranged by turns every two lattices, like a
checker-board, and $\dot\Phi$ is set to 0 for all lattices. The
initial string number per horizon volume in this case is as large as
$100$.  The result is depicted in Fig.\ \ref{fig:fitting} together
with the previous result. The number per horizon volume rapidly
decreases once and then rebounds and continues to increase, as in the
previous case. This result implies that the upward trend in the
previous simulation is genuine, because, if it were spurious, it would
gradually decrease and approach the asymptotic constant value from
above. In the next section we present a phenomenological analytic
model to explain why a departure from the scaling solution is
obtained.

\section{Analytic explanation}

\label{sec:model}

In the present simulations in 2+1 dimensions, strings are represented
by point-like quasi-particles and anti-particles depending on the
direction of the winding. Here, strings cannot intercommute with each
other to form loops, but rather they can only pair-annihilate upon
collision. Between a string and an anti-string works an attractive
force proportional to the inverse separation, which is a consequence
of the extended field gradient as a global defect. Seen from the
simulation result, as the system relaxes sufficiently and enters a
quasi-equilibrium state, a string usually remains at rest because it
loses kinetic energy due to cosmic expansion. Once within the horizon,
a pair of strings feel the above long-range force, and they start
moving toward each other.  Therefore, we have only to elucidate the
dynamics of a pair of particles obeying such a force in an expanding
universe.

Let $n(t)$ be the average number of strings per unit physical volume
at the cosmic time $t$. When the system is in a quasi-equilibrium
state, the probability per unit time $P(t)$ that a string annihilates
with an anti-string is written by \beq P(t) \sim \frac{1}{T(t)} \:,
\eeq

\noindent
where $T(t)$ is the period required for a pair of strings at rest with
mean separation, $\sim 1/n(t)$, to pair-annihilate. Since $n(t)$ is
diluted by cosmic expansion and the above annihilation process, it
satisfies the Boltzmann equation,
\bea
  \frac{dn(t)}{dt} &=& - P(t)\,n(t) - 2 H n(t) \:, \non \\
                   &=& - \frac{n(t)}{T(t)} -
                         \frac{n(t)}{t} \:.
\label{eqn:evolution}
\eea           

In order to determine the time dependence of $T(t)$, we investigate
the relative motion of particles subject to an attractive force
proportional to the inverse separation in an expanding universe, which
is described by the equation
\beq
  \frac{d^2r}{dt^2} + \frac{1}{t}\frac{dr}{dt} = -\frac{2\pi}{a(t)^2r}
  \:,
\eeq

\noindent
where $r$ is the comoving separation. In terms of the physical
separation $x \equiv a(t) r = (t/t_{i})^{1/2} r$, this equation reads
\beq
  \frac{d^2x}{dt^2} + \frac{1}{4}\frac{x}{t^2} = -\frac{2\pi}{x} \:.
\label{eqn:diff}
\eeq

\noindent
For the moment, we neglect the dissipation term due to cosmic
expansion. This is justified below. Then integration of Eq.\ 
(\ref{eqn:diff}) over $t$ gives
\beq
  \frac12 \lmk \frac{dx}{dt} \rmk^2 
        + 2\pi \ln \lmk \frac{x}{x_{0}} \rmk = 0 \:,
\eeq

\noindent
where $\frac{dx}{dt} = 0$ and $ x = x_{0}$ at $t = 0$. Then the period
$T$ is estimated as
\bea
  T &=& \int_{0}^{T} dt = \int_{x_{0}}^{d} 
      \frac{-dx}{\sqrt{-4\pi \ln \lmk \frac{x}{x_{0}} \rmk}} \non \\
    &=& \frac{1}{2\sqrt{\pi}}
        \lhk\, \lkk\,-2x \sqrt{\ln \lmk \frac{x_{0}}{x} \rmk}
                \,\rkk_{d}^{x_{0}}
        + 2 \int_{d}^{x_{0}} \sqrt{\ln \lmk \frac{x_{0}}{x} \rmk} dx
          \,\rhk \non \\  
    &\sim& \sqrt\frac{1}{\pi}\,x_{0}\,\sqrt{\ln \lmk \frac{x_{0}}{d} \rmk}
\:,
\label{eqn:period}
\eea

\noindent
where $d$ is the core radius. Before inserting this into Eq.\ 
(\ref{eqn:evolution}) to solve for $n(t)$, we justify the omission of
the dissipation term. A pair of strings at rest initially with
separation $x_{0}$ approach a separation $x$ in a time interval $t$
given by
\beq
  t \sim \sqrt\frac{1}{\pi}\,x_{0}\,\sqrt{\ln \lmk \frac{x_{0}}{x} \rmk} \:.
\eeq

\noindent
Then the dissipation term, $x/(4t^2)$, reads $\pi x /(4 x_{0}^2
\ln(x_{0}/{x}))$. Since $x$ is smaller than $x_{0}$, the term from the
potential force in Eq.\ (\ref{eqn:diff}), $2\pi/x$, always dominates
over the dissipation term.

Taking $x_{0}$ as the mean separation, $1/\sqrt{n(t)}$, Eq.\ 
(\ref{eqn:evolution}) becomes
\bea
  \frac{dn(t)}{dt} &=& - \frac{\sqrt{\pi}\,n(t)^{\frac32}}
                 {\sqrt{\ln \lmk \frac{1}{d\sqrt{n(t)}} \rmk}}  -
                         \frac{n(t)}{t} \non \\
                   &\sim& - \frac{\sqrt{\pi}\,n(t)^{\frac32}} 
                 {\sqrt{\ln \lmk \frac{A t}{d} \rmk}}  -
                         \frac{n(t)}{t} \:,
\eea 

\noindent
where $1/\sqrt{n(t)}$ is nearly proportional to the cosmic time from
the simulation result, and the proportional coefficient is $A$. 
~Substituting $f(t)$ for $n(t) t$, we have
\beq
  \frac{1}{t} \frac{df}{dt} = - \frac{\sqrt{\pi} f^{3/2}}
                             {\sqrt{\ln \lmk \frac{A t}{d} \rmk}}
                              \frac{1}{t^{3/2}}  \:.
\eeq

\noindent
Integrating over both sides gives
\beq
  \frac{1}{f^{1/2}} \sim \sqrt{\pi}\, B  \frac{t^{1/2}}
                             {\sqrt{\ln \lmk \frac{A t}{d} \rmk}}
                               + C  \:,
\eeq

\noindent
where $B$ represents a numerical factor $\sim \CO(1)$ and $C$ is an
integration constant determined from the initial conditions.

Then $n(t)$ is given by
\beq
  n(t) \sim \frac{D}{t^2}\, 
         \frac{ \ln{\lmk \frac{t}{E} \rmk} }
              { \lkk\,1 + \frac{C D^{1/2}}{t^{1/2}} 
                    \sqrt{\ln{\lmk \frac{t}{E} \rmk}} \,\rkk^{2}  } 
\:,
\eeq

\noindent
where $D = (\pi B^{2})^{-1}, E \sim d/A \sim d/\sqrt{D}$.
~Therefore the number of strings per horizon volume is given by
\bea
  n(t) H^{-2} &\sim& 4D\,
         \frac{ \ln{\lmk \frac{t}{E} \rmk} }
              { \lkk\,1 + \frac{C D^{1/2}}{t^{1/2}} 
                    \sqrt{\ln{\lmk \frac{t}{E} \rmk}} \,\rkk^{2}  }
  \non \\
  &\Longrightarrow& 4D\,
         \ln{\lmk \frac{t}{E} \rmk} \:.
   \qquad \qquad  (\mbox{asymptotically})
\eea

\noindent
Taking $D \sim 0.1$, $E \sim 0.4$~($d \sim 1.0$) and $C \sim -0.09$, we
can fit the results of the simulation by the above formula. This is
depicted in Fig.\ \ref{fig:fitting}. The number of strings per horizon
volume is not a constant, but proportional to $(\ln t)$ asymptotically.
Note that only the constant $C$ depends on the initial conditions.
Taking $C \sim 0.8$ with almost the same $D$ and $E$, that from the
checker-board initial conditions can also be fitted. This implies that
our fitting formula is consistent.

\section{Summary and Discussion}

\label{sec:summary}

In this paper, we have reported the dynamics of a complex scalar field
from the symmetric phase to the broken phase in 2+1 dimensions.
Starting from the thermal equilibrium state, we have observed how
strings are formed as the temperature decreases and the symmetry is
spontaneously broken. We also find that, as the string system grows,
it does not go into a scaling regime, where the number of strings per
horizon volume stays constant, but the number increases in proportion
to $(\ln t)$ asymptotically.  This can be interpreted as follows.  Our
simulation was performed in 2+1 dimensions and strings are in fact
point-like. Thus, the dominant energy loss process is not the formation
of loops through the intercommutation of infinite strings, but rather
pair-annihilation of straight strings. They move under logarithmic
potentials, which causes deviation from scale-invariance.

\section*{Acknowledgments}
MY is grateful to Professor K. Sato for his continuous encouragement
and Dr.\ T.\ Shiromizu and Professor M.\ Morikawa for their useful
comments. JY would like to thank Professor Andrei Linde for his
hospitality at Stanford University, where part of this work was done.
This work was partially supported by the Japanese Grant-in-Aid for
Scientific Research from the Monbusho, Nos.\ 10-04558~(MY), 09740334~(JY)
and 10640250~(MK). JY acknowledges support from the Monbusho.

\newpage

\begin{figure}[htb]
  \begin{center}
    \leavevmode\psfig{figure=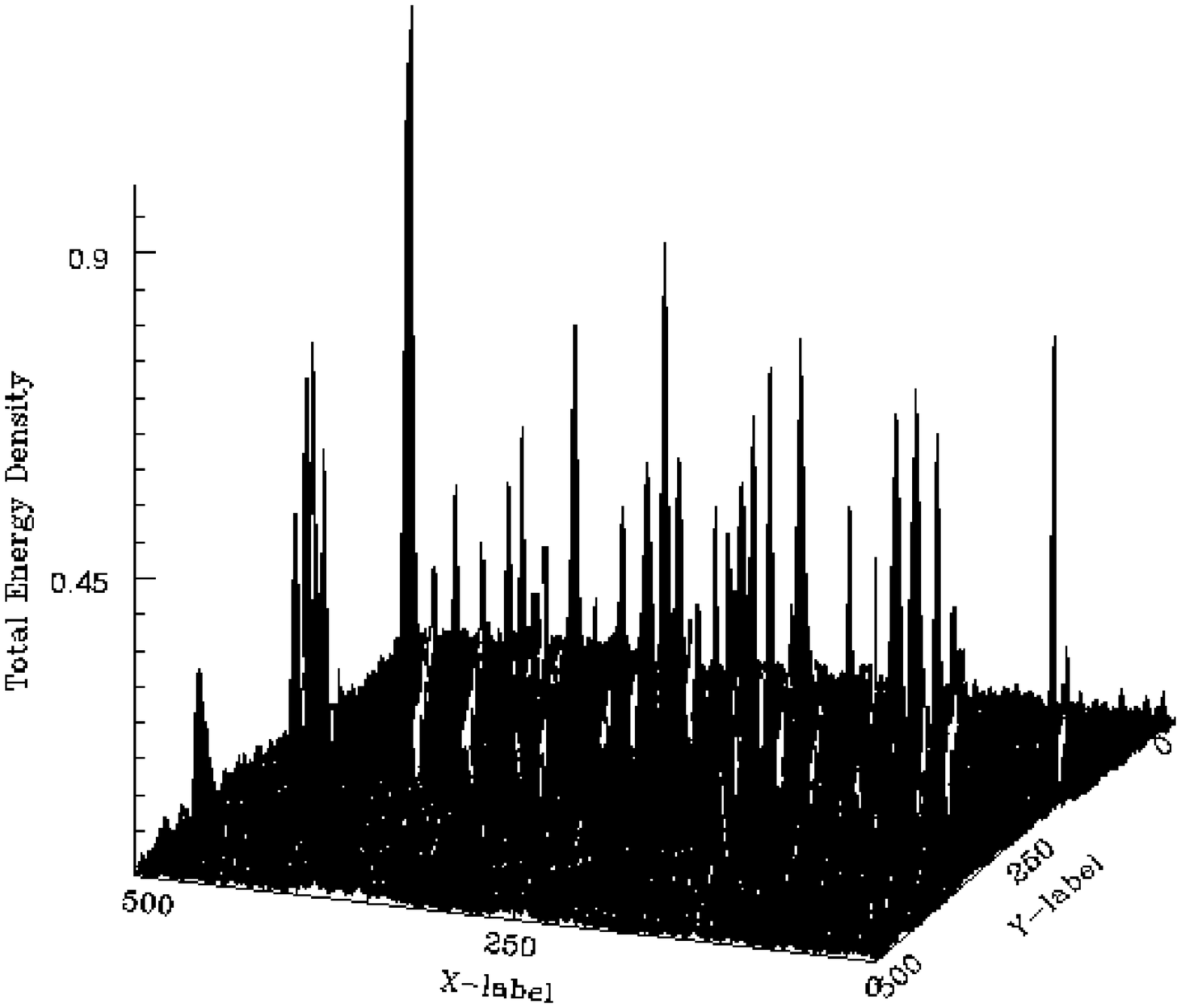,height=15cm}
  \end{center}
  \caption{That at $t = 50$.}
  \label{fig:evolution2}
\end{figure}

\begin{figure}[htb]
  \begin{center}
    \leavevmode\psfig{figure=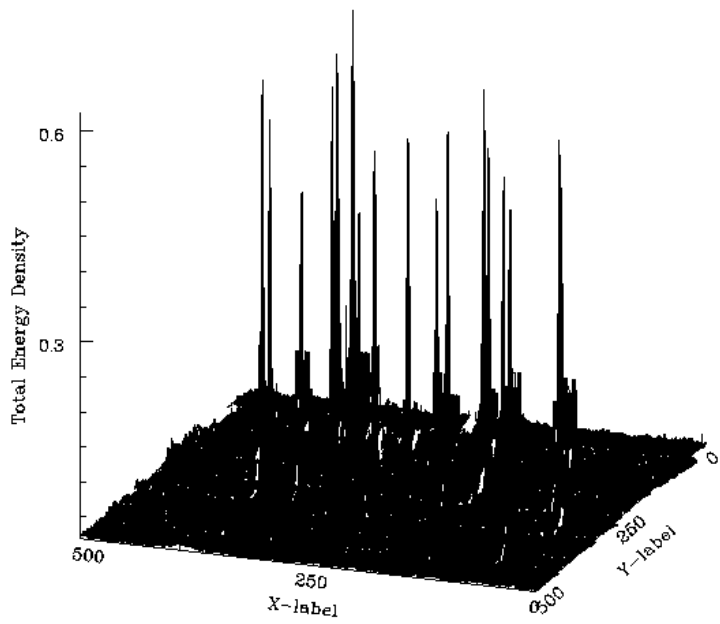,height=15cm}
  \end{center}
  \caption{That at $t = 100$.}
  \label{fig:evolution3}
\end{figure}

\end{document}